\documentclass[10pt]{article}

\usepackage{graphicx}
\usepackage{citesort}
\usepackage{equations}

\newcommand{\dotprod}{\,{\scriptscriptstyle \stackrel{\bullet}{{}}}\,}

\makeatletter

\def\compoundrel#1\over#2{\mathpalette\compoundreL{{#1}\over{#2}}}
\def\compoundreL#1#2{\compoundREL#1#2}
\def\compoundREL#1#2\over#3{\mathrel{\vcenter{\hbox{$\m@th\buildrel{#1#2}\over{#1#3}$}}}}
\makeatother

\begin{document}

\begin{center}
  \Large\bf Simulating Complex Dynamics In Intermediate
  And Large-Aspect-Ratio Convection Systems
\end{center}
\vspace{1ex}
\begin{center}
  Ming-Chih Lai\\
  Department of Mathematics, Chung Cheng
  University, Minghsiung, Chiayi 621, Taiwan\\
  \ \\
  Keng-Hwee Chiam and M.C. Cross\\
  Department of Physics, California Institute of
  Technology, Pasadena, CA 91125\\
  \ \\
  Henry Greenside\\
  Department of Physics, P. O. Box 90305, Duke
  University, Durham, NC 27708-0305
\end{center}
\vspace{3ex}
\begin{center}         
  \bf Abstract  
\end{center}

Buoyancy-induced (Rayleigh-B\'enard) convection of a
fluid between two horizontal plates is a central
paradigm for studying the transition to complex
spatiotemporal dynamics in sustained nonequilibrium
systems. To improve the analysis of experimental data
and the quantitative comparison of theory with
experiment, we have developed a three-dimensional
finite-difference code that can integrate the
three-dimensional Boussinesq equations (which govern
the evolution of the temperature, velocity, and
pressure fields associated with a convecting flow)
efficiently in large box-shaped domains with
experimentally appropriate lateral boundary conditions.
We discuss some details of this code and present two
applications, one to the occurrence of quasiperiodic
dynamics with as many as 5~incommensurate frequencies
in a moderate-aspect-ratio $10\times 5$ convection
cell, and one to the onset of spiral defect chaos in
square cells with aspect ratios varying from
$\Gamma=16$ to~56.


\vspace{1ex}

\section{Introduction}
\label{sec-intro}

A frontier of great importance for DOE-related research
is the study of sustained nonequilibrium dynamical
systems, for which imposed external fluxes of energy
and matter can lead to states that vary temporally and
spatially in a complex way~\cite{Cross93}. Despite the
collaboration of theorists, computational scientists,
and experimentalists over the last thirty years and
despite the great need to solve numerous practical
engineering problems, many basic questions about
sustained nonequilibrium states remain unanswered.
Researchers would like to know what possible states can
occur for specified external fluxes, how to predict
when one state will change into another as some
parameter is varied, how transport of energy and matter
depends on the spatiotemporal structure of a state, and
whether one can select particular states by appropriate
external perturbations so as to optimize a system for a
particular goal. While experiments and simulations have
been useful in suggesting what possibilities can occur
(e.g., the surprising experimental discovery of the
spiral defect chaos state~\cite{Morris93}, a numerical
example of which is shown below in
Fig.~\ref{Fig-Spirals}), there remains a great need to
develop a stronger theoretical and conceptual
foundation that can unify the many observations and
that can improve both experimental and computational
investigations.

Perhaps the simplest and best idealized experimental
system for exploring basic questions and principles of
nonequilibrium systems is Rayleigh-B\'enard convection,
which has become an experimental and theoretical
paradigm for many researchers~\cite{Cross93}. A
Rayleigh-B\'enard experiment consists of a thin layer
of fluid confined between two horizontal
spatially-uniform constant-temperature metal plates
such that the bottom plate is maintained at a constant
higher temperature than the upper plate. As the
temperature difference (or its dimensionless
equivalent, the Rayleigh number~R) is increased in
successive constant steps, the fluid first makes a
transition from a motionless structureless state to
cellular overturning convection rolls and then to ever
more complex dynamical states which eventually become
nonperiodic in space and time. Convection has
significant advantages over other experimental systems
in having static homogeneous boundary conditions, in
having no net flow of fluid through the system, in
allowing precise and reproducible experiments with good
visualization, and in being amenable to a quantitative
mathematical description through the so-called
Boussinesq equations.

In this paper, we report applications of a new computer
code to two intriguing convection experiments. The code
is the first of several being developed and applied by
a Caltech-Duke collaboration whose long-term goal is to
understand convection phenomena more quantitatively,
especially in the large-aspect-ratio limit (cells whose
widths are large compared to their depths) which
experiments have shown to be of great interest even
close to the onset of convection, where analytical
progress is most likely to be possible~\cite{Cross93}.
Our code differs from some other recently developed
codes~\cite{Decker94SDC} primarily through the
inclusion of experimentally appropriate lateral
boundary conditions (rather than periodic boundary
conditions) on the velocity and temperature fields so
that the forcing due to lateral boundaries can be taken
into account. In the following sections, we give a
brief summary of the code followed by a discussion and
demonstration of how the code can provide new insights
about two poorly understood experimental phenomena, the
occurrence of dynamics with many incommensurate
frequencies in a moderate-aspect ratio convection cell
first observed by Walden et al~\cite{Walden84} and the
onset of spiral defect chaos in domains of varying size
(which has not yet been studied experimentally). These
results provide new and detailed examples of the
substantial influence of lateral boundaries on
nonequilibrium dynamics.

\section{Numerical Integration Of The 3D Boussinesq
Equations}
\label{sec-integration}

Since technical details of our numerical algorithm will
be available elsewhere~\cite{Lai00}, we provide only
some motivation for and highlights of our numerical
method. The goal is to integrate the five coupled
three-dimensional nonlinear partial differential
equations known as the Boussinesq equations which state
(under certain assumptions not given here) the local
conservation of momentum, energy, and mass for parcels
of fluid subjected to buoyancy forces. By scaling time,
space, and field magnitudes in appropriate ways, one
can write the Boussinesq equations in the following
dimensionless form:
\begin{eqnarray}
  \partial_t{\bf u} &=&
    - {\bf u} \dotprod \nabla {\bf u} 
    - \nabla p 
    +  \sigma \nabla^2{\bf u} 
    + \sigma {\rm R} T \hat{z}
    , \label{eq:momentum-eq} \\
  \partial_t{T} &=&
  - {\bf u} \dotprod \nabla T
  + \nabla^2 T ,
  \label{eq:temp-eq} \\
  \nabla \dotprod {\bf u} & = & 0
  , \label{eq-divu} 
\end{eqnarray}
where~${\bf u}(t,{\bf x}) = \left( u_x(t,{\bf
    x}),u_y(t,{\bf x}),u_z(t,{\bf x}) \right)$ is the
velocity field at time~$t$ and position~${\bf
  x}=(x,y,z)$, $T(t,{\bf x})$ is the temperature field,
$p(t,{\bf x})$ is the pressure field, $\sigma$ is the
fluid's Prandtl number which is assumed to be
independent of temperature and so a constant, and~R is
the Rayleigh number which is the key parameter that is
varied in most experiments and simulations, usually
with all other parameters held fixed. In this paper, we
study these equations in a simple box geometry of
dimensions~$\Gamma_x \times \Gamma_y \times 1$; the
quantities~$\Gamma_x$ and~$\Gamma_y$ are ratios of
lateral widths to the unit fluid depth and are called
aspect ratios.  Since the fluid is confined by
stationary material walls, the velocity vanishes at
these walls which provides the following boundary
condition on~$\bf u$:
\begin{equation}
  {\bf u} = 0 \qquad \mbox{on all walls} .
\end{equation}
With our rescaled variables, the constant temperature
boundary conditions on the bottom and top plates ($z=0$
and~$z=1$ respectively) are simply
\begin{equation}
  \label{eq:vert-temp-bcs}
  T(t,x,y,0) = 1
   \qquad \mbox{and} \qquad
  T(t,x,y,1) = 0 .
\end{equation}
The temperature field~$T$ satisfies an additional
boundary condition on the lateral walls which, for this
paper, we take to be a no-flux condition corresponding
to a perfect thermal insulator
\begin{equation}
  \label{eq:lateral-temp-bc}
  \partial_n T = \hat{\bf n} \dotprod \nabla T = 0
  \qquad 
  \mbox{on lateral walls,}
\end{equation}
where~$\hat{\bf n}$ is the normal unit vector at a
given point on the wall. However, the code is more
general and can treat thermal boundary conditions that
interpolate between conducting and insulating
sidewalls.

The numerical challenge is to integrate these equations
and boundary conditions efficiently and accurately over
long time intervals in large cells of simple geometry;
boxes and cylinders cover nearly all the experimental
cases while a box with periodic sidewalls is useful for
comparing with theory.  Rayleigh-B\'enard convection is
so important that many numerical methods have been
developed and tried over the years although, somewhat
unfortunately, most of these methods have not been
compared with each other to determine which best
achieves a practical balance of efficiency, accuracy,
ease of programming, and parallel scalability on some
specific computer architecture.  Because our interest
is to study fundamental questions in simple cell
geometries, we chose not to use finite element or
spectral element methods whose main strengths are the
ability to handle irregular boundaries.  Because our
short term needs are for modest accuracy, simplicity
and flexibility of coding, and good parallel scaling on
Beowulf-style computers, we chose second-order-accurate
finite-difference approximations on Cartesian meshes
instead of spectral methods.

Our code uses a traditional time-splitting method in
which higher-order linear operators are advanced
implicitly in time and lower-order nonlinear terms are
advanced explicitly~\cite{Lai00}, achieving at each
time step an overall accuracy of second order in time.
The incompressibility condition $\nabla\dotprod{\bf
  u}=0$ is treated by a standard projection
method~\cite{Bell89} in which the momentum conservation
equations are used to update the current velocity~${\bf
  u}(t,{\bf x})$ to an intermediate field~${\bf
  u}^\ast$ that is not divergence free, and then~${\bf
  u}^\ast$ is ``projected'' onto a divergence-free
field~${\bf u}(t+\Delta{t},{\bf x})$ by solving a
Poisson equation for the new pressure field.  To
advance one time step~$\Delta{t}$, four 3D~Helmholtz
equations and one 3D~Poisson equation must be solved
with appropriate boundary conditions, and the solution
of these linear equations constitute the most time
consuming part of the code.  For this first generation
code, we used FISHPACK fast direct solvers (available
through www.netlib.org) which are well suited for
modest-aspect-ratio problems on single-processor Alpha
workstations. Future codes will use parallel iterative
methods which are also better suited for the
non-constant-coefficient linear operators that arise in
a cylindrical geometry.

Our code was innovative mainly through the use of {\em
  colocated} meshes, in which all field values and all
operators of field values were evaluated on the same
set of mesh points.  For two- and three-dimensional
fluid simulations of incompressible flow, empirical
studies and some analysis have suggested that staggered
meshes (in which scalar quantities are stored at the
centers of grid boxes while vector components are
stored on the faces or vertices of the boxes) were
necessary to avoid numerical instabilities associated
with the pressure~\cite{Harlow65}.  Our colocated-mesh
Boussinesq code proved to be numerically stable which
led to a substantial reduction in the effort of writing
and validating the code compared to a staggered-mesh
code.  For lack of space, we refer to our forthcoming
paper for further details, e.g., how our code was
validated and its efficiency and accuracy as a function
of various parameters~\cite{Lai00}.

\section{Applications}
\label{sec-applications}

We now report on two preliminary applications of the
above convection code. First we try to simulate an
intriguing experiment~\cite{Walden84} that goes to the
heart of how chaotic behavior arises in a continuous
medium, here through the occurrence of quasiperiodic
states with as many as five incommensurate frequencies.
The mystery to understand is the spatial structure of
the different oscillations and their dependence on
aspect ratio and Rayleigh number. Second, we
investigate how the onset of spiral defect chaos
state~\cite{Morris93} depends on the aspect
ratio~$\Gamma$ of a square box, which we increase in
small successive increments.  Varying the aspect ratio
is difficult in laboratory experiments and these
calculations demonstrate the usefulness of having
quantitatively accurate codes to complement
experiments.

\subsection{Multi-Frequency Dynamics at Intermediate
Aspect Ratios}
\label{sec-many-frequencies}

Our first calculation was motivated by the experimental
paper of Walden et al~\cite{Walden84}, which reported
in~1984 the unexpected occurrence of spatiotemporal
quasiperiodic states in a convecting flow with as many
as five incommensurate frequencies. This result seemed
to contradict one of the major mathematical insights of
the time, a theorem of Newhouse, Ruelle, and
Takens~\cite{Ott93} which argued that chaotic behavior
should be typically observed after at most three
successive Hopf bifurcations since quasiperiodic
dynamics with three or more incommensurate frequencies
can be perturbed infinitesimally to become chaotic.
Although the abstract mathematical arguments were
difficult to interpret for laboratory experiments and
despite clarifications of this theorem by later
numerical simulations on simple map
systems~\cite{Grebogi85}, it is still not understood
how a physical continuous medium can develop so many
independent oscillations or whether a physical
mechanism can be identified for each independent
frequency.

To make contact with this experiment, we have carried
out the first (to our knowledge) simulations in a
box-like domain with parameters nearly identical to
those of the experiment. Thus we performed numerical
integrations of the 3D~Boussinesq equations in a cell
of aspect ratio~$9.5 \times 4.5 \times 1$, for a fluid
with Prandtl number~$\sigma = 3.5$ (corresponding to
water with a mean temperature of~$50^\circ\rm C$) and
over a comparable range of Rayleigh numbers up to $\rm
R = 20 R_c$, where~$\rm R_c \approx 1708$ is the
critical value for the onset of convection in an
infinite-aspect-ratio cell.  The most poorly justified
approximation was our choice of laterally insulating
sidewalls Eq.~(\ref{eq:lateral-temp-bc}) since the real
experiment had finitely conducting glass sidewalls
between a copper bottom plate and sapphire upper plate.
(The thermal diffusivities~$\kappa$ of copper, glass,
sapphire, and water are respectively 1.20, 0.004,
0.113, and~0.00147 $\rm cm^2/sec$.)  A typical run used
a resolution of $76\times 36 \times 8$ points and a
constant time step of~$\Delta{t} = 0.001$.  A run to
collect 65,000 points took approximately 1.5~hours on a
Compaq XP1000~workstation using a 667~MHz 264~Alpha
chip with a 4~MB cache.

\begin{figure}
  [tbh]
\begin{center}
  \includegraphics[ width=6in]{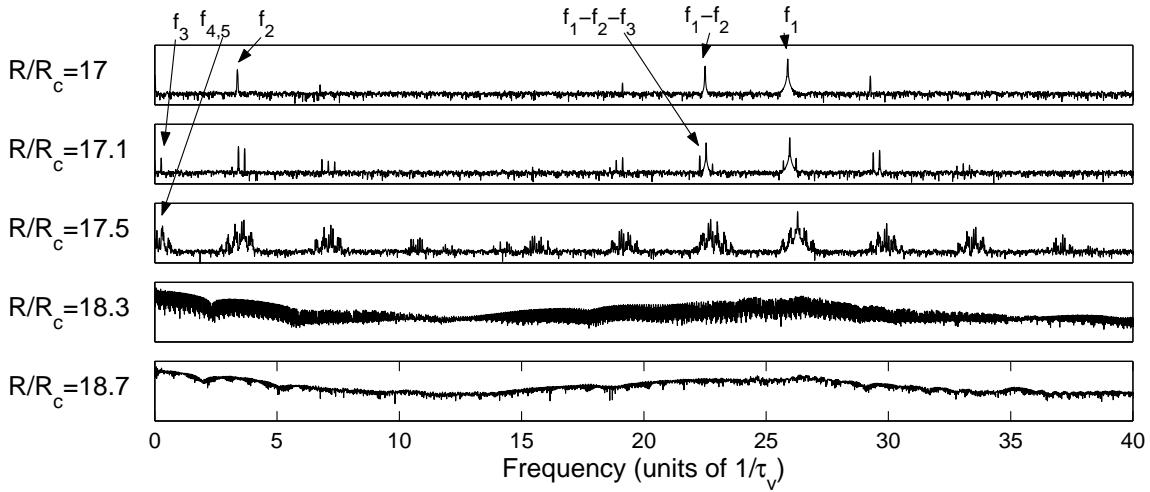}
\caption{
  Power spectra~$P(f)$ versus frequency~$f$ for five
  values of the Rayleigh number~R over the range $\rm
  17 \le R/R_c \le 18.7$ in a cell of aspect ratio
  $9.5\times 4.5\times 1$ and for a fluid of Prandtl
  number~$\sigma=3.5$.  The top three panels show
  quasiperiodic motion with~2, 3, and~5 incommensurate
  frequencies respectively.  The last two panels show
  spectra of chaotic dynamics with continuous
  broad-band features.}
  \label{Fig-PowerSpectra-R}
\end{center}
\end{figure}

Some representative results are shown in
Fig.~\ref{Fig-PowerSpectra-R}.  As the Rayleigh
number~R is increased in small steps, new
incommensurate frequencies appear until, at $\rm
R/R_c=17.5$, 5~incommensurate frequencies are observed
just as in the experiment. The fact that these
frequencies were incommensurate was supported by
plotting (not shown) the ratios of frequencies
corresponding to different peaks and observing that
these ratios varied smoothly with~R, i.e., no mode
locking to a rational value took place.
The numerical simulations did not reproduce
quantitatively the magnitude of the lower frequencies
observed in experiment. For example, for the
4-frequency convection state, the simulation has a low
frequency peak at~$f_4 \approx 0.17$ which is about a
factor of three smaller than that observed in the
experiment. A first guess is that this discrepancy is a
consequence of the convenient but experimentally
inaccurate no-heat-flux boundary condition
Eq.~(\ref{eq:lateral-temp-bc}).

\begin{figure}[tbh]
\begin{center}
\includegraphics[ width=5.0in ]{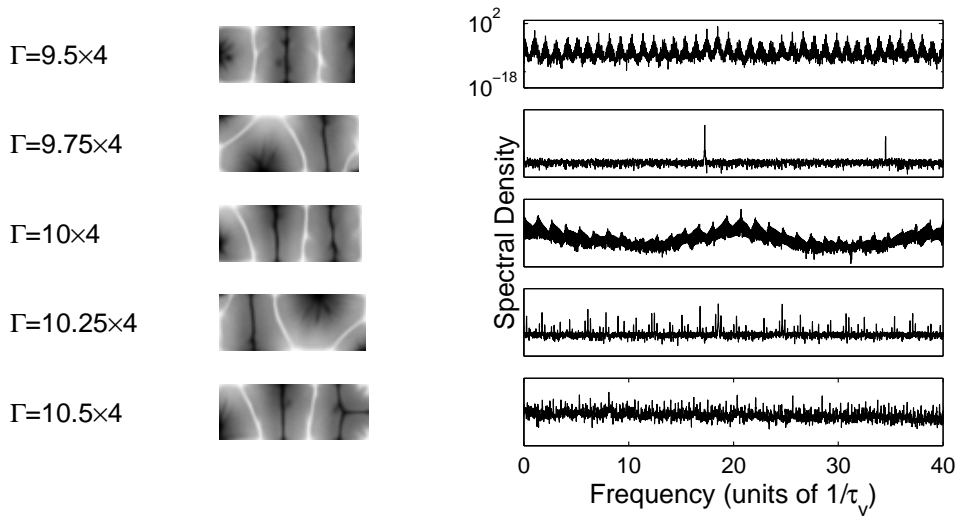}
\end{center}
\caption{
  Changes in convection dynamics as the aspect
  ratio~$\Gamma_x$ is increased in small steps for
  fixed~$\Gamma_y = 4$ and $\rm R = 18 R_c$. The left
  column of plots are instantaneous density plots of
  the temperature field at the midplane~$z=1/2$ with
  light regions corresponding to warm fluid, dark
  regions to cool fluid. The right column of plots are
  corresponding power spectra~$P(f)$ calculated from
  time series of 65,536 values of the temperature at
  the midpoint of the cell. Rows~1, 3, and~5 are
  chaotic, row~2 is periodic, and row~4 is
  quasiperiodic with three independent frequencies.}
  \label{Fig-VaryGamma}
\end{figure}

In related simulations, we have also explored how the
dynamics depended on aspect ratio, a question which is
difficult to explore experimentally.
Fig.~\ref{Fig-VaryGamma} shows several instantaneous
convection patterns and the power spectra of the
corresponding time-dependent states over the range~$9.5
\le \Gamma_x \le 10.5$ with~$\Gamma_y$ and~R held
fixed.  A surprising and new result is that small
changes in~$\Gamma$ lead to dramatically different
patterns and dynamics.  Indeed, for the states of
Fig.~\ref{Fig-VaryGamma} and others not shown over this
same range, one can identify time independent,
periodic, quasiperiodic (with~3 and~4 frequencies), and
chaotic dynamics. The spatiotemporal dynamics is
evidently highly sensitive to small changes in the
system geometry at these intermediate aspect ratios.

\subsection{Onset of Spiral Chaos}
\label{sec-spiral-chaos}

As the technology improved for exploring
large-aspect-ratio convection dynamics,
experimentalists made a remarkable discovery
in~1993~\cite{Morris93} of an intricate spatiotemporal
chaotic convecting flow in a cylindrical geometry near
onset, a regime that previous experiments in smaller
aspect ratios had suggested would show only simple
convective patterns, and for which theory predicts that
parallel time-independent convection rolls should be
stable~\cite{Cakmur97}.  This {\em spiral defect chaos}
state (so named because of the unexpected occurrence of
rotating spiral structures) remains poorly understood
seven years later and is now regarded by many
convection researchers to be an especially important
example of spatiotemporal chaos to understand.
Intriguing and also poorly understood is the
experimental observation that spiral defect chaos is
observed only when the aspect ratio~$\Gamma$ of the
cylindrical cell is sufficiently large, with the radius
being at least 40~times the fluid depth.

\begin{figure}[ptbh]
\begin{center}
\includegraphics[width=5.8in] {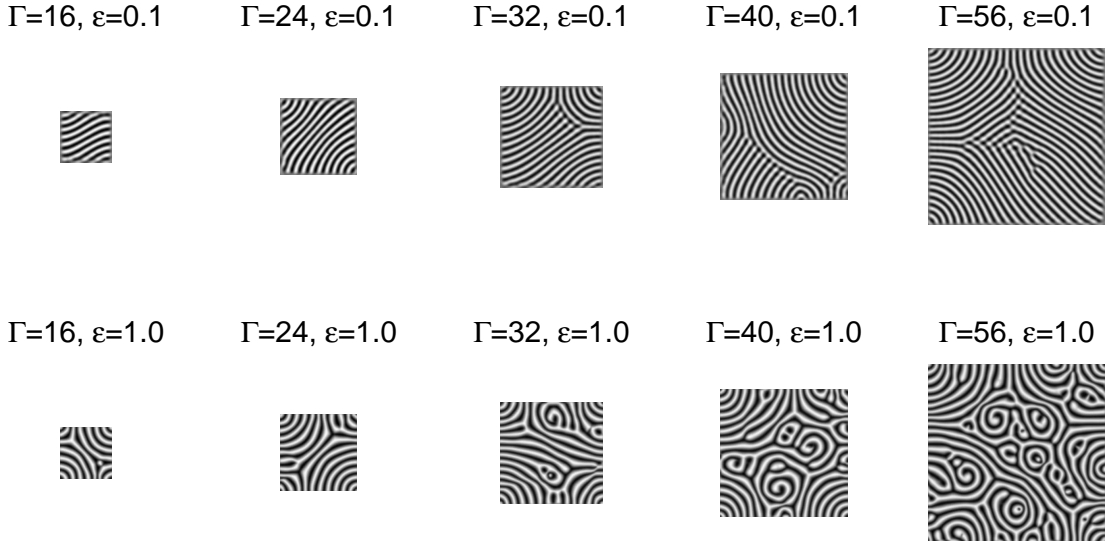}
\end{center}
\caption{
  Instantaneous patterns observed in various aspect
  ratios for two values of the reduced Rayleigh number
  $\varepsilon=(R-R_{c})/R_{c}$ and for a fixed Prandtl
  number of $\sigma = 0.96$, corresponding to the
  compressed CO$_{2}$ gas used in the experiments. The
  first two columns of states are time independent. }
  \label{Fig-Spirals}
\end{figure}

Using the code described above, we have explored for
the onset and properties of spiral defect chaos in
finite cells with experimentally realistic lateral
boundary conditions and with varying aspect ratio,
although for a square rather than cylindrical geometry.
Representative results for two different values of the
reduced Rayleigh number~$\varepsilon= \rm (R -
R_c)/R_c$ are shown in~Fig.~\ref{Fig-Spirals}.
For~$\Gamma \le 24$, the asymptotic dynamics are
stationary while time-dependent states are observed for
larger~$\Gamma$, with spirals being observed only for
the larger Rayleigh numbers. Spirals appear in square
geometries for smaller aspect ratios than those of a
cylindrical cell at the same reduced Rayleigh number.

\begin{figure}
[ptbh]
\begin{center}
  \includegraphics[width=4.5in]{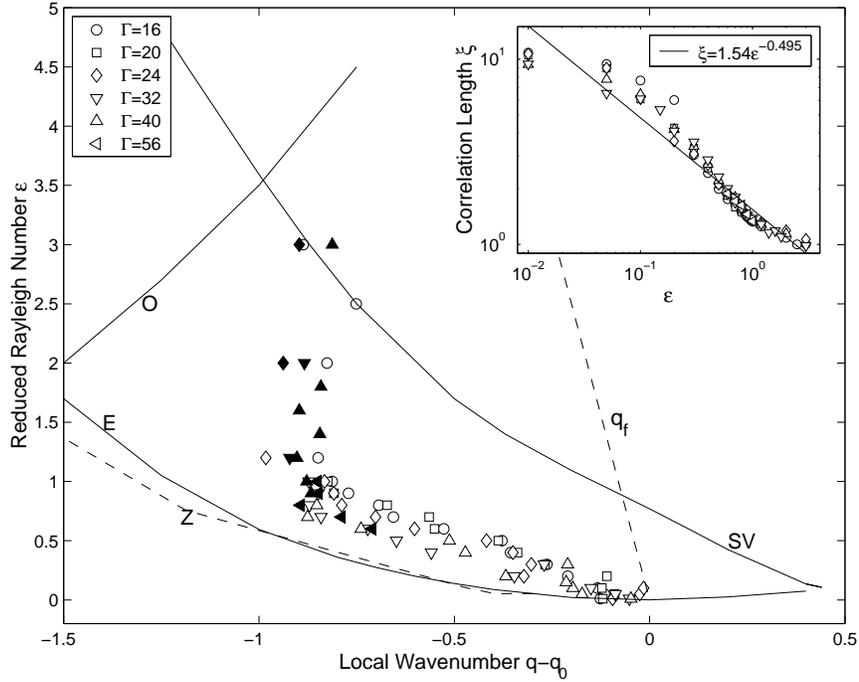}
  \caption{
    Plot of the deviation $\bar{q}-q_0$ of the mean
    wave number~$\bar{q}$ from the critical wave number
    $q_0=3.117$ as a function of the reduced Rayleigh
    number~$\varepsilon$. Also shown are the
    instability boundaries~\cite{Busse78} which limit
    the range for the ideal roll state in a laterally
    infinite geometry (SV=skew varicose; O=oscillatory;
    E=Eckhaus; Z=zigzag), and the wave number
    $q_f(\varepsilon)$ that is selected in axisymmetric rolls.
    Solid symbols denote states where dynamic spiral
    defects are observed. The inset shows the
    correlation length $\xi$ defined as the inverse of
    the width of the wave number probability
    distribution $P(q)$. The straight line has a
    slope~$1/2$ as would be predicted by the amplitude
    equation theory near threshold.}
  \label{Fig-PhaseDiagram}
\end{center}
\end{figure}

As a first step towards quantifying and analyzing these
complex patterns, we have calculated the time-averaged
distribution~$P(q)$ of local wave numbers~$q$ as a
function of aspect ratio and Rayleigh number. Following
a recent suggestion of Egolf et al~\cite{Egolf98sdc},
we estimated local wave numbers~$q(t,x,y)$ from the
ratio~$-\nabla^2{\theta}/\theta$ where
$\theta=\theta(t,x,y,1/2)$ is the deviation of the
temperature field~T from its linear conducting profile,
evaluated at the cell midplane~$z=1/2$. The
distribution $P(q)$ was then obtained by averaging many
instantaneous histograms of~$q$ over time.  A
compilation of the mean wave numbers~$\bar{q}$
associated with each wave number distribution is shown
in Fig.~\ref{Fig-PhaseDiagram}, which shows rather
remarkably that the trend for the variation
of~$\bar{q}$ with~R is nearly independent of the aspect
ratio, and that~$\bar{q}$ decreases roughly linearly
with increasing Rayleigh number up to $\rm R/R_c \simeq
2$.  Near the value $q-q_0=-0.8$ (with~$q_0$ the
critical wave number at onset), there is a dramatic
change with $\bar{q}$ becoming essentially independent
of~R.  At this point, spiral defect chaos develops in
the larger aspect ratio cells while the smaller cells
are chaotic and lack spiral defects.

The average spatial disorder of each pattern can also
be quantified by a correlation length~$\xi$, which is
defined here to be the inverse of the width of the
distribution~$P(q)$. The inset in
Fig.~\ref{Fig-PhaseDiagram} shows that~$\xi$ is also
insensitive to the aspect ratio and obeys approximately
a power law dependence~$\varepsilon^{-1/2}$ which is
the same as that predicted by the amplitude equation
theory~\cite{Cross93} (although the range of Rayleigh
numbers in the plot is much larger than the range over
which this theory might be expected to hold).  A
similar trend has again been noticed in cylindrical
geometry experiments, although an experiment in a
rectangular cell found a divergence at a nonzero value
of $\varepsilon$.
%

\begin{figure}[ptbh]
\begin{center}
\includegraphics[ width=5.5in ]{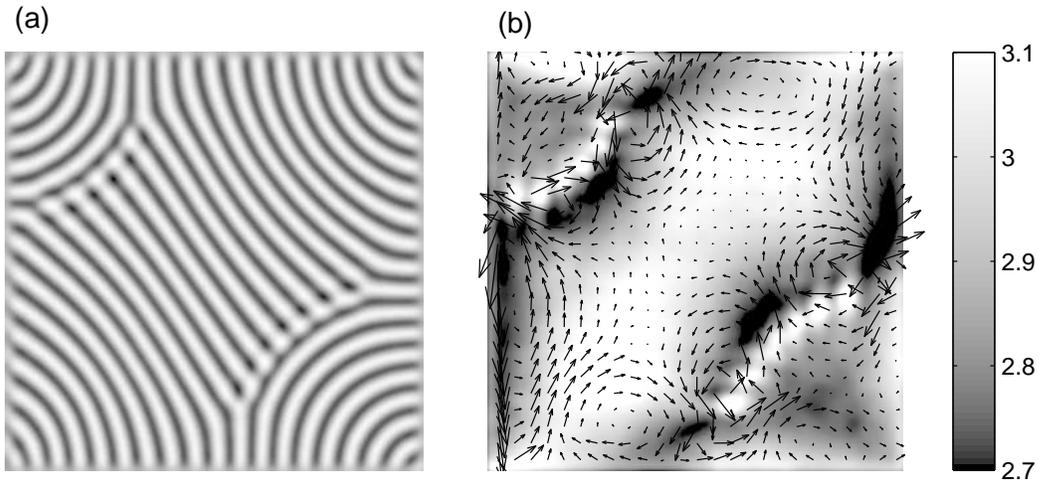}
\end{center}
\caption{
  (a) Roll pattern and (b) wave number distribution
  (gray scale) at $R/R_{c}=1.15$ in a cell of aspect
  ratio $\Gamma=40$. The corresponding mean flow
  (arrows) was estimated by integrating the horizontal
  velocity components of~$\bf u$ vertically over the
  fluid depth. The maximum Euclidean norm of~$\bf u$
  has the value~6.1 in units of $d/t_v$ while the mean
  flow is much smaller, with a corresponding maximum
  magnitude of~0.12.}
  \label{Fig-MeanFlow}
\end{figure}

The trend of $\bar{q}(\varepsilon)$ in
Fig.~\ref{Fig-PhaseDiagram}, which has also been
observed in cylindrical experiments, is far from that
predicted theoretically by Cross and
Newell~\cite{Cross84}, who argued that portions of
circular rolls around a ``focus singularity'' such as
those spanning the corners in the cells of
Fig.~\ref{Fig-Spirals} should lead to the selection of
the same wave number $q_f$ as that selected in
concentric axisymmetric rolls~\cite{Buell86pf}.  The
discrepancy is particularly striking in the simple
structures seen at low values of $\varepsilon$ such as
Fig.~\ref{Fig-MeanFlow}, where theory~\cite{Cross84}
suggests that the focus singularities in the corners
should determine the wave number over
much of the system.
One possible way in which arcs of rolls can act
differently than complete circles is that arcs can
drive a ``mean flow'', which may then modify the wave
number distribution. The mean flow is, roughly, the
horizontal fluid flow integrated across the depth of
the cell and cannot occur in axisymmetric (or straight)
roll configurations because of the incompressibility of
the fluid. The mean flow is known to be important in
producing the skew-varicose instability and in
suppressing the zigzag instability for Prandtl numbers
of order unity.

Using the detailed knowledge provided by the code of
the convection pattern, the wave number distribution,
and the mean flow, we are able to assess for the first
time the importance of the mean flow in producing the
deviations of the measured wave number $\bar{q}$ from
$q_f$.  Fig.~\ref{Fig-MeanFlow} shows the distribution
of the wave number field and corresponding mean flow.
In regions towards the center of the cell where the
mean flow is small, the wave number is indeed close to
the predicted value $q_f=3.1$.  However circulating
mean flow patterns develop in the cell corners and the
wave numbers there are substantially reduced below
$q_{f}$. A plot of the distribution
$P(q)$ (not shown) indicates in fact that the
\emph{largest}~$q$ with significant probability
is close to $q_{f}$ but the spread of $P(q)$
to smaller values of $q$ means that the \emph{mean}
$\bar{q}$ is considerably below $q_{f}$.  Given the characteristic
form of the mean flows that form in the
corners---consisting of two regions of counter-rotating
vorticity---an analytic attack on this long standing
question is an appropriate next step.

\section{Conclusions}
\label{sec-conclusions}

As initial applications of our intermediate- to
large-aspect-ratio fluid convection code we have
studied two aspects of the onset of chaotic dynamics.
In both of these examples the role of the physical
boundaries were found to play a vital role---in the
intermediate aspect ratio by determining the basic
structures about which dynamics develops, and in the
large aspect ratio cell where the mean flows that form
in the corners of the cell play an important roll in
determining the wave number distribution---and so the
physical issues are not accessible to previous codes
where periodic boundary conditions are used. The
preliminary results we present here suggest further
directions to explore, both numerically and
analytically.

\section{Acknowledgements}

This work was supported by the Engineering Research
Program of the Office of Basic Energy Sciences at the
Department of Energy, Grant DE-FT02-98ER14892. We also
thank Paul Kolodner for useful discussions concerning
the convection apparatus of Walden et al.

\bibliographystyle{unsrt}
\bibliography{cfd,chaos,convection,control,na,spatiotemporal,stc}
\label{bibliography}

\end{document}